\documentclass[twocolumn,showpacs,preprintnumbers,amsmath,amssymb]{revtex4}
\usepackage{graphicx}% Include figure files
\usepackage{dcolumn}% Align table columns on decimal point
\usepackage{bm}% bold math

\begin{document}

\title{Tachyonic potential in Bianchi type-I universe }
\author{P.K.Suresh}
\affiliation{
School of Physics,University of Hyderabad.Hyderabad 500 046.India.}
 \email{pkssp@uohyd.ernet.in}

\date{\today}

\begin{abstract}

  Motivated form recent string theoretic results, a tachyonic potential is constructed for a spatially homogeneous and anisotropic background cosmology. 

\end{abstract}

\pacs{98.80.Cq}

\maketitle

\section{Introduction}

Particles with velocities greater than the speed of light, popularly known as 
tachyon, has been received much attention in cosmology  recently 
\cite{1,2}. Though the subject of tachyon is still speculative, its 
various properties has been studied \cite{3}. Recently, Sen \cite{4} has 
shown that a rolling tachyonic condensate can be studied with the 
help of the energy density and pressure of an effective fluid. 
Following these lines, Finstein \cite{5} has obtained an exact 
solution for a spatially flat isotropic universe in terms of 
tachyonic potential. Using the tachyonic condensate, Padmanabhan 
\cite{6}has studied the possibility of expanding universe driven by 
tachyonic matter. Cosmological inflation driven by the rolling tachyon 
in the context of the brane world  scenario is also investigated \cite{7,8}.
Sami has constructed tachyonic potential which can implements power-law inflation in the brane world cosmology\cite{9}. The possibility of tachyonic inflation with exponential also studied\cite{11}.
These studies indicate, the evolution of the 
tachyon condensate can have cosmological significance and may be 
useful to understand various cosmological issues. The 
aforementioned tachyonic condensate has explored in the context 
of isotropic Friedman-Robertson-Walker (FRW) cosmological model.

The observed large scale isotropy of the universe have been established for times after the era at which the universe become transparent to radiation, and can successfully account by FRW model. This does not mean that the extrapolation to earlier times the model is equally suitable to describe the very early unverse, especially near the Planck scale or string scale. There exist a wide class of 
anisotropic cosmological models, which also often studying in 
cosmology, due to various reasons \cite{11}. 
 There are theoretical arguments that sustain the existence of an anisotropic phase that approaches an isotropic case\cite{12}. Also, anisotropic cosmological models  are found a suitable candidate to avoid the assumption of specific initial conditions in FRW models. The early universe could also  characterized by irregular expansion  mechanism. Therefor,  it would be useful to explore  cosmological models in which anisotropic, existing at  early stage of expansion, are damped out in the course of evolution. Interest in such models have been received much attention since 1978\cite{13}.    
 Among the anisotropic 
models, the Bianchi type - I cosmological model is the simplest 
one, which is an anisotropic generalization of the Friedman model 
with Euclidean spatial geometry, and that contains special isotropic cases and allows arbitrary small anisotropy levels \cite{14}. Since present universe is isotropic, it makes an appropriate candidate for studying the possible effects of  anisotropy in the early universe on present-day observations. There is evidence that the dynamics of the early universe may have been profoundly influenced by the presence of spatial anisotropie just below the Planck or string scale \cite{15}. Also the solution of  the low-energy string cosmological effective action are by their nature anisotropic\cite{16,17}. Therefor it would be useful to study the role of tachyonic field  in Bianchi type -I cosmological model.

Usually, scalar field potential is used to drive the expansion of 
the universe. Since there exist a mapping between  tachyonic and 
scalar field potentials directly \cite{6}, it is reasonable to expect 
that tachyonic potential can also be used to drive the expansion 
of the universe. Thus, to study the expansion of the universe, 
driven by tachyonic source, it is instructive to determine the 
tachyonic potential in a specific cosmological model under 
consideration. The present work is to obtain the tachyonic 
potential in Bianchi type -I unverse.
 
For the present work we take $c=1$.

\section{Perfect fluid and scalar field in Bianchi type-I  metric }

Consider spatially homogeneous, anisotropic, and topologically 
three - torus or Euclidean  background metric 
\begin{equation}\label{1}
dS^{2}= -dt^{2}+ \sum _{i=1}^{3} R_{i}^{2}(t)\left(dx^{i} \right)^{2},
\end{equation}
where $ R_{1}, R_{2}, $ and $  R_{3} $ are the scale factors in 
$x,y,$ and $ z$ directions respectively. The metric (\ref{1}) is an 
anisotropic generalization of the Friedman model with Euclidean 
spatial geometry. The three scale factors  are determined vis 
Einstein equations.

Next, we consider certain procedure to construct some useful relations and can be used  to obtain potential for scalar field  and tachyonic field in Bianchi type-I spacetime.
For this, assume that the spacetime (\ref{1}) filled with a perfect fluid.
Investigations of Bianchi type-I  spacetime filled with a perfect fluid  has studied  by many authors \cite{18,19,20}.
The energy -momentum tensor for the perfect fluid can be written as
\begin{equation}\label{2}
   T_{\mu \nu }= -p g_{\mu \nu }+ (\varrho+p)v^{\mu} v_{\nu},
\end{equation} 
 where $p$ and $\rho$ are respectively known as  pressure and energy density for the perfect fluid.
Thus for  metric (\ref{1}) Friedman equations  becomes:
\begin{equation}
\frac{ \dot{R}_{1} \dot{R}_{2}} { R_{1} R_{2} } + \frac{ 
\dot{R}_{2} \dot{R}_{3}} { R_{2} R_{3} }+ \frac{ \dot{R}_{1} 
\dot{R}_{3}} { R_{1} R_{3} }= 8 \pi G \rho ,
\end{equation}
\begin{eqnarray}
\nonumber-2 \left(\frac{\ddot{R}_{1}}{R_{1}}+ \frac{\ddot{R}_{2}}{R_{2}} 
+ \frac{\ddot{R}_{3}}{R_{3}}\right)-
\frac{\ddot{R}_{1}R_{2}}{R_{1}R_{2}} \\
-\frac{\ddot{R}_{2}R_{3}}{R_{2}R_{3}}-
\frac{\ddot{R}_{1}R_{3}}{R_{1}R_{3}} 
=24 \pi p G,  
\end{eqnarray}
Thus
\begin{equation}\label{3}
\sum _{i=1}^{3}\frac{{\ddot{R_{i}}}}{R_{i}}= - 4\pi G\left(\rho + 
3p \right).
\end{equation}

It is convenient to define a parameter
\begin{eqnarray}
\omega (t)\equiv  \frac {p \left(t \right)}{\rho \left(t 
\right)}.
\end{eqnarray}

The equation of motion for the present situation can be written as
\begin{eqnarray}
 d\left( \rho R_{1} R_{2} R_{3}\right)= - \omega \rho d\left( R_{1} R_{2} R_{3}\right),
\end{eqnarray}
which leads to
\begin{eqnarray}
\frac{{\dot{\rho}} }{\rho }= - \sum _{i= 1}^{3} H_{i}\left(t 
\right) \left(\omega \left(t \right) + 1\right).
\end{eqnarray}
In the anisotropic background metric (\ref{1}), one can write $ \rho 
\propto (H_{1}H_{2}+ H_{2}H_{3} +H_{1}H_{3})$, so that
\begin{eqnarray}
\frac{\dot{\rho }}{\rho 
}= \frac{\dot{\cal{H}}}{\cal{H}},
\end{eqnarray}
where
\begin{eqnarray}\label{4}
{\cal{H}}=H_{1}H_{2}+H_{2}H_{3}+H_{1}H_{3}.
\end{eqnarray}
and
\begin{eqnarray}
\dot{\cal{H}}&=& \dot{H_{1}}H_{2}+\dot{H_{2}}H_{1}+\dot{H_{2}}H_{3}+ \\ \nonumber
&& \dot{H_{3}}H_{2}+ 
\dot{H_{1}}H_{3}+\dot{H_{3}}H_{1} .
\end{eqnarray}

Therefor
\begin{eqnarray}\label{6}
\omega + 1= 
- \frac{1}{\sum _{i= 1}^{3} H_{i}}\frac{\dot{\cal{H}}}{\cal{H}}.
\end{eqnarray}

The procedure  of computing the above relation will be useful for further study.

Since there exist a mapping between the scalar field potential 
and tachyonic potential \cite{6}, it would be useful to compare  both 
potentials in Bianchi type-I cosmological model. Thus, we first 
obtain the potential for the scalar field. 

 Consider a $k=0$ 
universe with a scalar field as the source  and its   Lagrangian 
is given by
\begin{equation}
L= \frac{1}{2}\partial _{\alpha }\varphi\partial ^{\alpha }\varphi- U_{\varphi}(\varphi).
\end{equation}

Therefor  the equation of motion for the scalar field for the metric (\ref{1})becomes
\begin{equation}
\ddot{\varphi} + \sum_{i=1}^{3} H_{i}\dot{\varphi}= - \frac{dU_{\varphi}} {d \varphi}.
\end{equation}

Assume that the evolution of the universe is already specified  
so that $ R_{i}$, (i=1,2,3), $H_{i}= \frac { \dot{R_{i}}} {R_{i} 
}$.. etc, are known functions of time and we need to determine 
$U_{\varphi} $, 
Since the motion under consideration is spatially homogeneous,i.e., $ 
\varphi (t,x)= \varphi (t) $, the energy density and pressure of 
the scalar field are respectively given by
\begin{equation}\label{5}
\rho _{\varphi}= \frac{1}{2} \dot{\varphi }+ U_{\varphi}; 
\hspace{1cm} p_{\varphi}= \frac{1}{2} \dot{\varphi } - U_{\varphi}.
\end{equation}

From the definition of $ \omega $ and  (\ref{5}), it follows that
\begin{equation}
\frac{\dot{\varphi}^2}{U_{\varphi}}= \frac{\left( 1+ \omega_{\varphi} \right)}{\left( 
1- \omega_{\varphi} \right)}\equiv Q(t) ,
\end{equation}
say. Thus $ {\dot{\varphi}}^2 = 2 Q U_{\varphi}$, differentiating with 
respect to time and using (\ref{6})  we get
\begin{equation}
\frac{\dot{U_{\varphi}}}{U_{\varphi}}= - \frac{\dot{Q} + 2 Q \sum_{i=1}^{3} 
H_{i}}{1+Q}.
\end{equation}
Integrating this expression and using the definition of $Q(t)$ and  adopt the method of obtaining (\ref{6})  for the scalar field, we find
\begin{eqnarray}
U_{\varphi}(t)=\frac{\cal{H} }
{8 \pi 
G}  \left(1+  
\frac{1}{2}\frac{\dot{\cal{H}}} 
{\sum_{i=1}^{3}H_{i}(\cal{H})}\right).
\end{eqnarray}
Substituting back in the expression $ {\dot{\varphi}}^2 = 2 Q U_{\varphi}$, 
we find
{\small
\begin{equation}
\varphi(t)= \int dt \sqrt{- \frac{\mathcal{H}}{4 \pi 
G}\frac{\left(\frac{\dot{\mathcal{H}}}{\sum_{i=1}^{3}H_{i}\mathcal{H}}\right)} 
{\left(2+\frac{\dot{\mathcal{H}}}{\sum_{i=1}^{3}H_{i}\mathcal{H}}\right)} 
\left(1+\frac{\dot{\mathcal{H}}}{2\sum_{i=1}^{3}H_{i}\mathcal{H}}\right)}
\end{equation}}
where $\mathcal{H}$ is given by (10).  Eqs (18) 
and (19) completely solve the task of obtaining a potential $ 
U_{\varphi}$, which will lead to a given set of $ R_{1}, R_{2}$, 
and $R_{3}$. These equations determine $\varphi(t)$ and $U_{\varphi}(t)$ in 
terms of  $ R_{1}, R_{2}$, and $R_{3}$ thereby implicitly 
determine $U_{\varphi}$.

\section{Tachyon field in Bianchi type-I metric}

Our next aim is to obtain the tachyonic potential in the Bianchi 
type - I universe. The action for tachyonic field coupled with Einstein gravity in the low energy limit is given by \cite{21}

\begin{eqnarray}\label{20}
S= \int d^{4}x\sqrt{-g}\left( \frac{R}{16 \pi G}- U_{\top}(\top) \sqrt{1+\partial_{\mu} \top \partial^{\nu}\top}\right).
\end{eqnarray}  
Where $g $ is the determinant of the metric $g_{\mu \nu}$ and $R$ its scalar curvature. 
 Since  (\ref{20}) has a potential function 
$U_{\top} $, one could expect that 
 cosmological evolution can be obtained with tachyonic 
 field as the source by an approximate potential taken in a convenient form.

The tachyon field under our consideration is assumed as spatially homogeneous,
i.e,$ \top(x,t)=\top(t)$, then
its 
 energy density $\rho _{\top}$ and pressure 
$p_{\top} $ respectively  are given by [4] 
\begin{eqnarray}\label{21}
    \rho_{\top} = \frac{U_{\top}} { \sqrt{1- {\dot{\top}}^2} }.
\end{eqnarray}
and
\begin{eqnarray}\label{22}
p_{\top}=-U_{\top} \sqrt{1- {\dot{\top}}^2}.
\end{eqnarray}
Consider a universe with $k=0$ for the spacetime (\ref{1}) filled with  a spatially homogeneous tachyon field.
The equation of motion of tachyon field in the spacetime (\ref{1}) takes following form
\begin{eqnarray}
\frac{\ddot{\top}}{1-\dot{\top}^{2}} + \sum_{i=1}^{3} \frac{\dot{R}_{i}}{R_{i}} \dot{\top}+ \frac{U_{,\top}}{U_{\top}}=0.
\end{eqnarray}  

For the tachyonic field with (\ref{21}) and (\ref{22}), we get
\begin{equation}
\omega_{\top}+1= {\dot{\top}}^{2}.
\end{equation}
Again  the method of obtaining (12) can be used to compute  $ \omega_{\top}$ and applying the result in (24), we get
\begin{equation}
{\dot{\top}}^{2}=  - 
\frac{\dot{\mathcal{{H}}}}{{\mathcal{H}}\sum_{i=1}^{3}H_{i}}.
\end{equation}
Thus it follows
\begin{equation}
{\top}= \int dt \sqrt{ - 
\frac{\dot{\mathcal{{H}}}}{{\mathcal{H}}\sum_{i=1}^{3}H_{i}} }.
\end{equation}

The Friedmann equation for the tachyon field  can be written  by using (\ref{21}) as
\begin{equation}
\frac{ \dot{R}_{1} \dot{R}_{2}} { R_{1} R_{2} } + \frac{ 
\dot{R}_{2} \dot{R}_{3}} { R_{2} R_{3} }+ \frac{ \dot{R}_{1} 
\dot{R}_{3}} { R_{1} R_{3} }= 8 \pi G \rho_{\top}.
\end{equation}

Thus, we get
\begin{eqnarray}
   U_{\top}    &= &   \frac{\cal{H}  }{8 \pi G} 
 \left(  1+  
\frac{\dot{\cal{H}}} {{\cal{H} }\sum_{i=1}^{3}H_{i}}
 \right)^{1/2}\\ \nonumber
 &   =&\left(- \omega_{\top}  \right) ^{1/2} \rho_{\top} .       
\end{eqnarray}
Equations (26) and (28) completely solve the problem of determine 
$U_{\top}$ and $\top  (t)$, for a given set of $ R_{1}$, $R_{2}$ 
and $R_{3}$.

Consider a universe with power-law expansion $ 
R_{1}=t^{n_{1}},R_{2}=t^{n_{2}}, $ and $ R_{3}=t^{n_{3}}$, thus, 
from (25), the complete solution can be obtained as
\begin{eqnarray}
\top = \left(\frac{2}{n_{1}+ n_{2}+ n_{3}} \right)^{1/2} t+ 
\top_{0},
\end{eqnarray}
and
\begin{equation}
U_{\top}=\left( \frac{n_{1}n_{2}+ n_{2}n_{3}+n_{1}n_{3}} {8 \pi 
G}\right) \left(1- \frac{2}{n_{1}+ n_{2}+ n_{3}} \right)^{1/2}
\frac{1}{t^{2}}.
\end{equation}
Combining  (29) and (30), we find the potential is
\begin{eqnarray}
U_{\top}&=& \frac{ 1} {4 \pi G }\left(  \frac{n_{1}n_{2}+ 
n_{2}n_{3}+n_{1}n_{3}} {n_{1}+n_{2} +n_{3}}\right)\\
&&\times \left(1- 
\frac{2}{n_{1}+ n_{2}+ n_{3}} \right)^{1/2}  \frac{1}{(\top - 
\top_{0})^{2}}. \nonumber
\end{eqnarray}
Thus
\begin{equation}
U_{\top}\propto \frac{1}{\top^{2}}.
\end{equation}

The rapid expansion is possible for the potential (31) with large 
values of $n_{i}$, as in the case of isotropic FRW model. Also 
the potential has reasonable behavior of $U_{\top}\rightarrow 
0$ as $\top\rightarrow \infty$ though its form for small and 
intermediate values of $\top$ is not supported by string 
theory. Comparing the potentials between the scalar field (18) 
and tachyonic field (28) in Bianchi type - I background metric, 
it shows that the  forms of the both potentials exhibit similar 
behavior.

To see the asymptotic form of the evolution of the solutions 
found above, assume that the late time behavior of $\top $ is 
given by
\begin{eqnarray}
\nonumber\top&=&\left(\frac{2}{n_{1}+n_{2}+n_{3}}\right)t+ \alpha 
e^{-\beta 
t} \\ 
&=&\left(\frac{2}{n_{1}+n_{2}+n_{3}}\right)t+{\mathcal{O}}(e^{-\beta 
t })
\end{eqnarray} 
where $n_{1},n_{2},n_{3},\alpha$ and $\beta$ are constants with 
2$\leq n_{1}+n_{2}+n_{3}$.

The value of $n_{1}+n_{2}+n_{3}$ is restricted to 2 because 
$\top(t)$ to grow proportional to $t$ asymptotically with 
exponentially small corrections. Thus the asymptotically (late 
times) behavior of the potential can be written as
\begin{eqnarray}
U_{\top}&\approx &\frac{ 1} {4 \pi G }\left(  \frac{n_{1}n_{2}+ 
n_{2}n_{3}+n_{1}n_{3}} {n_{1}+n_{2} +n_{3}}\right) \\ 
\nonumber && \times \left(1- 
\frac{2}{n_{1}+ n_{2}+ n_{3}} \right)^{1/2}  \top^{-2} \left[ 1 
+ {\mathcal{O}}(\frac{e^{-\beta t }}{t})\right]. \nonumber
\end{eqnarray}

 If we restrict the total value of $ n_{1},n_{2}$ and $n_{3}$, 
such that
\begin{equation}
\sum_{i=1}^{3}n_{i}\leq 2,
\end{equation}
then the potential can have values depending on the values of 
$n_{i}$'s. The potential give non-zero value only whenever two n 
are non-zeros.
 When $ R_{1}= R_{2}= R_{3}= R $ and $n_{1}= n_{2}= n_{3} = 
n$, then the result (30) is consistent with the isotropic case 
\cite{6}. 

\section{Conclusions }

In this brief work, we considered a tachyonic field as the source 
of gravity in the Bianchi type-I background metric and obtained 
the tachyonic potential. It is noted that for a large values of 
$n_{i}$'s, the potential behaves power-law expansion as in the 
case of isotropic FRW cosmological model. 
For a given set of  scale factors, $ R(t)_{i}$,
 it is always possible to obtain a potential, $U_{\top}$ 
 such that it result in a cosmological evolution. 
 Thus the tachyonic  potentials can be also used as a possible candidate to drive the expansion of  anisotropically expanding  universe, described by the metric(\ref{1}). 
The asymptotic nature of 
the potential is also discussed. We also obtained potential of scalar field in the Bianchi type-I cosmological model. 
The present study can account 
for the tachyonic potential for the Bianchi type-I cosmological 
model and is found consistent with isotropic  FRW cosmological 
model when $n_{1}$=$n_{2}$=$n_{3}$=$n$. \\

\begin{acknowledgments}
I thank Prof.Bindu A Bambah for useful discussion.
 I wish to thank the Director,IUCAA,Pune for warm hospitality and library facilities extended to me and acknowledge the Associateship of IUCAA.
\end{acknowledgments}

\end{document}